\documentstyle[12pt]{article}
\begin{document}
\onecolumn

\begin{titlepage}
\hfill gr-qc/0003098
\begin{center}
{\LARGE Solutions to Cosmological Problems with Energy Conservation 
and Varying c, G and $\Lambda$}\\  
\vspace{.5in}
P. Gopakumar and G.V. Vijayagovindan\\ 
School of Pure and Applied Physics, \\
Mahatma Gandhi University, \\ 
Kottayam - 686 560,\\
India.\\
\vspace{.5in}
\end{center}
\begin{abstract}
	The flatness and cosmological constant problems are solved
	with varying speed of light c, gravitational coupling strength 
	G and cosmological parameter $\Lambda$, by explicitly assuming
	energy conservation of observed matter. The present solution to
	the flatness problem is the same as a previous solution in
	which energy conservation was absent.\\
\end{abstract}	
\end{titlepage}

As an alternative to the inflationary model of the universe, varying speed of 
light theories $^{\cite{Moffat, CM, AM,Barrow,BM}}$ had been introduced.
Experimental observation of the 
variation of fine structure constant with time has been indicated by  
quasar absorption spectra$^{\cite{Webb}}$. Variations of fine structure constant
could 
be interpreted as variation of the speed of light or of the fundamental 
charge, e. In the model of Moffat$^{\cite{Moffat}}$, variation of speed of light
arises due 
to the spontaneous breakdown of local Lorentz invariance in the early universe.
In a later model$^{\cite{CM}}$, he introduced a dynamical mechanism of varying
speed of 
light (VSL) by working in a bi- metric theory. 
Kristsis$^{\cite{Kirt,Alex}}$ has given a VSL theory in 3+1 dimensions by
starting from a 
string theory motivated theory of branes.  Albrecht and Magueijo$^{\cite{AM}}$
and 
Barrow$^{\cite{Barrow}}$ 
 consider not only models with VSL but also those allowing G and $\Lambda$
 to vary with respect
to time in the conservation equations.

In this paper we consider a VSL theory in which the energy-momentum of 
 matter is conserved. We have reformulated the solutions to the 
cosmological problems on this basis.\\

\section{Flatness Problem}

\hspace{.2in} Albrecht and Magueijo proposed that a time varying speed of light c 
should not introduce 
changes in the curvature terms in the Einstein's equations in the cosmological 
frame and that Einstein's equations must still hold. Assuming that matter 
behaves as a perfect fluid, the equations of state can be written as, 
\begin{eqnarray}\label{1}
  P=(\gamma -1)\rho c^{2}(t).
\end{eqnarray}
Friedmann equations for a homogeneous space time, with c and G, as functions of 
time are, 
  \begin{eqnarray}\label{2}
  \frac{\dot{a}^{2}}{a^{2}}=\frac{8\pi}{3}G(t)\rho - \frac{Kc^{2}(t)}{a^{2}}
\end{eqnarray}  
\begin{eqnarray}\label{3}
  \ddot{a}= -\frac{4}{3}\pi G(t)(\rho+\frac{3p}{c^{2}(t)}) a
\end{eqnarray}
where $\rho$ and p are density and pressure of the matter, and K is the 
metric curvature parameter.
Combining Eq.(2) and Eq.(3), the generalized conservation equation is, 
\begin{eqnarray}\label{4}
\dot{\rho}+3\frac{\dot{a}}{a}(\rho+\frac{p}{c^{2}})=-\rho\frac{\dot{G}}{G}+
\frac{3Kc\dot{c}}{4{\pi}G a^{2}}
\end{eqnarray}\label{}

We assume the conservation of ordinary matter; ie., the left hand side of Eq.(4)  
is zero. Thus the variations in c and G are such that the right hand side of
Eq.(4) 
is identically zero. 
Assuming $\rho\propto a^{-3\gamma}$ and 
$c(t)=c_{0}a^{n}$,where $c_{0}$ and n are constants.
The solution of the right hand side of Eq.(4), for $3\gamma +2n-2 \neq0$ is 
\begin{equation}\label{}
G = \frac{3K(c_{0})^{2}n}{4\pi\rho_{0}}\frac{a^{3\gamma+2n-2}}{3\gamma+2n-2}+B
\end{equation}
and for, $3\gamma+2n=2$ is
\begin{equation}
G = \frac{3K(c_{0})^{2}n}{4\pi\rho_{0}}\ln a+B
\end{equation}   
where B is a constant of integration. 
Thus the Friedmann equation for the $3\gamma +2n \neq2 $ case becomes 
\begin{eqnarray}\label{5}
\frac{\dot{a}^{2}}{a^{2}}=B'a^{-3\gamma}+\frac{K(c_{0})^{2}a^{2n-2}
(2-3\gamma)}{(3\gamma+2n-2)}
\end{eqnarray}
where B'is a constant.
The curvature term in Eq.(2) vanishes as the scale factor evolves,
if $\ddot{a}>0$ ie., $\rho+\frac{3p}{c^{2}}<0$ .
Also the Eq.(4) gives the solution, 
\begin{center} $\rho \propto a^{-3\gamma}$ for $\ddot{G}=\ddot{c}=0$
if $\rho+\frac{p}{c^{2}}\geq0$
\end{center}
Using the equation of state, Eq.(1), these conditions imply
\begin{center}
$0\leq\gamma<\frac{2}{3}$
\end{center}
The scale factor evolves as
\begin{center}
 $a(t)\propto t^{\frac{2}{3}\gamma}$  if $\gamma>0$
\end{center}
and\begin{center} $a(t){\propto} \exp{(H_{0}(t)}$     if $\gamma=0$
\end{center}
For $\gamma<\frac{2}{3}$, the requirement $\rho+\frac{3p}{c^{2}}<0$ implies
$p<-\frac{1}{3}{\rho}c^{2}$, ie., the curvature term will vanish at large a 
only if the matter stress is gravitationally repulsive.  
From Eq.(7) it can be seen that the flatness problem can be solved for,
\begin{eqnarray}
n \leq \frac{1}{2}{(2-3\gamma)}.
\end{eqnarray}
This is exactly the same inequality, which Barrow derived without assuming 
energy conservation of ordinary matter$^{\cite{Barrow}}$.

\section{Lambda Problem}

\hspace{.2in} To incorporate the cosmological constant term into the Friedmann equation, a 
vacuum stress is considered obeying the equation of state, 
\begin{eqnarray}
  p_{\Lambda}=-\rho_{\Lambda}c^{2} ,
\end{eqnarray}
 with 
\begin{eqnarray}
\rho_{\Lambda}=\frac{{\Lambda} c^{2}}{8{\pi}G}{\geq}0 .
\end{eqnarray} 
The Friedmann equation containing ${\rho}_\Lambda $is,
\begin{eqnarray}
\frac{\dot{a}^{2}}{a^{2}}=\frac{8}{3}{\pi}G({\rho}+{\rho}_{\Lambda})
-\frac{K c^{2}}{a^{2}} .
\end{eqnarray}
As the universe expands the term containing ${\rho}_\Lambda $ should dominate.
But observationally it is very small. This is the $\Lambda$ problem. 
The conservation Eq.(4) generalised to include ${\rho}_\Lambda $ is, 
\begin{eqnarray}
\dot{\rho}+3 \frac{\dot{a}}{a}[{\rho}+\frac{p}{c^{2}}]=-{\dot{{\rho}_{\Lambda}}}
-({\rho}+{\rho}_{\Lambda})
{\frac{\dot{G}}{G}}+{\frac{3Kc{\dot{c}}}{4{\pi}Ga^{2}}} .
\end{eqnarray}
We assume a form for the variation of G and $\Lambda$  
in terms of the scale factor  as,
$G=G_{0}a^{q}$ and $\Lambda=\Lambda_{0}a^{s}$ 
where $\Lambda_{0}$,q, $G_{0}$ and s are constants.
Now Eq.(12) can be written as, 
\begin{eqnarray}
{\dot{\rho}}+3{\frac{\dot{a}}{a}}{\frac{p}{c^{2}}}+{\frac{\dot{\Lambda}c^{2}}
{8{\pi}G}}=
-(\rho+\rho_{\Lambda})\frac{\dot{G}}{G}+\frac{3Kc\dot{c}}{4{\pi}Gc^{2}}-
\frac{2\Lambda c \dot{c}}{8 \pi G}+\frac{\Lambda c^{2}\dot{G}}{8 \pi G^{2}}
\end{eqnarray}
Assuming conservation of ordinary matter, we put the left hand side and 
right hand side of Eq.(13) separately to zero. The underlying assumption 
being that the variations of $\Lambda$, G and c are such that conservation of
energy, 
as true with the non varying parameters, is still valid. 
The solution for left hand side of Eq.(13), being zero, is 
\begin{eqnarray}
\rho_{\Lambda}=-\frac{\Lambda_{0}s(c_{0})^{2}}{8 \pi G_{0}}\frac{a(2n+s-q)}
{(2n+3\gamma+s-q)}+Ba^{-3\gamma}
\end{eqnarray}
where B is a constant of integration.
The right hand side of Eq.(13) can be written as, 
\begin{eqnarray}
\frac{\Lambda_{0}(c_{0})^{2}}{8 \pi G_{0}}[\frac{sq}{2n+3\gamma+s-q}-2n]
a^{2n+s-q-1}+\frac{3K(c_{0})^{2}n}{4 \pi G_{0}}a^{2n-q-3}
-qBa^{-3\gamma-1}=0
\end{eqnarray}
A solution for this equation is  
\begin{eqnarray}
  \frac{sq}{2n+3\gamma+s-q}=2n ,
\end{eqnarray}
\begin{eqnarray}
2n-q-3=-3\gamma-1 ,
\end{eqnarray}
and
\begin{eqnarray}
\frac{3K(c_{0})^{2}n}{4 \pi G_{0}}=qB .
\end{eqnarray}
Using Eq.(17), Eq.(16) can be written as, 
\begin{eqnarray}
\frac{qs}{s+2}=2n .
\end{eqnarray}
Then   
$q=\frac{3K(c_{0})^{2}n}{4 \pi G_{0}B} \equiv q_{0}n $ and
  $s=\frac{4}{q_{0}-2}$ .\\

For the dust era ($\gamma=1$) Eq.(17) becomes 
\begin{eqnarray}
n=\frac{1}{q_{0}-2} \hspace{.1in} or \hspace{.1in} 4n=s
\end{eqnarray}
 For n to be negative, 
 as in the solution of the flatness problem, $q_{0}<2$.
For the radiation dominated era ($\gamma=\frac{4}{3}$), Eq.(17) is 
\begin{eqnarray}
n=\frac{2}{q_{0}-2} \hspace{.1in} or\hspace{.1in} 2n=s
\end{eqnarray}

The Friedmann equation for a time varying cosmological parameter, Eq.(13),
 becomes, 
\begin{eqnarray}
\frac{\dot {a}^{2}}{a^{2}}=\frac{\Lambda_{0}(c_{0})^{2}}{3}
\frac{(2n+3\gamma-q)}{(2n+3\gamma+s-q)}
a^{2n+s}+\frac{8}{3} \pi BG_{0}a^{q-3\gamma}-K(c_{0})^{2}a^{2n-2}
\end{eqnarray}

For the dust dominated universe, $\gamma=1$ , using Eq.(20) 
\begin{eqnarray}
\frac{\dot{a}^{2}}{a^{2}}=\frac{2\Lambda_{0}(c_{0})^{2}}{3(4n+2)}a^{6n}
+\frac{8 \pi B G_{0}}{3}a^{2n-2}-K(c_{0})^{2}a^{2n-2}
\end{eqnarray}
For  $n<-\frac{1}{2}$ the cosmological term will go to zero at large times
 faster than the other two terms of Eq.(23).   
For $n<1$ the curvature term will tend to zero at large times. Thus for 
$n<-\frac{1}{2}$ all the terms  will vanish at large times solving
 both the cosmological and the flatness problems.

For the radiation dominated universe ($\gamma=\frac{4}{3}$ ), using equation
(21) 
\begin{eqnarray}
 \frac{\dot{a}^{2}}{a^{2}}=\frac{2\Lambda_{0}(c_{0})^{2}}{3(2n+2)}a^{4n}+
\frac{8 \pi B G_{0}}{3} a^{2n-2}-K(c_{0})^{2}a^{2n-2}
\end{eqnarray}
For $n<-1$ the $\Lambda$ term will go to zero faster than the other two terms
of the Eq.(24). As in the dust universe $n<1$ makes the curvature term vanish
at large times. Thus in the radiation dominated universe $n<-1$ solves both the
cosmological and flatness problems.

We have solved the flatness and $\Lambda$ problems in a Friedmann universe 
by assuming that the variation of c, G and $\Lambda$  are such that the
conservation of matter 
with fixed c, G and $\Lambda$  still holds. In the absence of $\Lambda$ the
solution is the same 
as that without assuming energy conservation. With $\Lambda$  we can still solve
the 
cosmological problem but with different exponents. Thus it might be worth 
exploring a more fundamental theory that will allow the variation of the 
parameters c, G, $\Lambda$  without violating energy conservation.

\section*{Acknowledgements}
PG thanks CSIR, NewDelhi for a research fellowship, and GVV thanks Prof. 
N. Dadhich for useful discussions and IUCAA, Pune for allowing to 
use their facilities.

\end{document}